# Dissecting the dot-com bubble in the 1990s NASDAQ

Yuchao Fan

University of Cambridge

Fanyuchao01@gmail.com

Abstract: In this paper, I revisit Phillips, Wu and Yu's seminal 2011 paper on testing for the dot-com bubble. I apply recent advancements of their methods to individual Nasdaq stocks and use a novel specification for fundamentals. To address a divide in the literature, I generate a detailed sectoral breakdown of the dot-com bubble. I find that it comprised multiple overlapping episodes of exuberance and that there were indeed two starting dates for internet exuberance.

## 1. Introduction

This paper examines the structure of exuberance in stocks traded on the Nasdaq stock exchange during the 1990s dot-com bubble. I build upon the seminal 2011 paper by Phillips, Wu and Yu (PWY) but apply new advances in time series methods as well as innovations in data and model specification.

My paper is the first to test for bubbles in individual Nasdaq stocks during the dot-com bubble. I achieve this by constructing a large dataset that covers the majority of stocks listed on the Nasdaq at the time. I ensure my tests are robust by considering a novel fundamental specification which is widely implemented by practitioners, but previously unseen within the bubble literature. I reconcile this specification with conventional economic theory and provide a detailed description of potential issues and solutions. Additionally, by testing individual stocks, I am able to provide a detailed sectoral breakdown of exuberance during the dot-com bubble, which to my knowledge has never been done before. In doing so, I am able to address a divide in the literature regarding when the dot-com bubble first began, and also offer an insight into whether it was rational or not.

## 2. Background, Literature Review and Motivation

Tim Berners-Lee's creation of the World Wide Web in 1990 heralded the beginning of the modern internet era. Internet usage increased meteorically across the US population, from 0.785% in 1990 to 43% by 2000 (Source: World Bank).

The proliferation of internet usage led to the formation of many internet-based companies, whilst existing companies tried to incorporate the internet into their operations. The dot-com bubble epithet is a direct reflection on how many companies at the time appended the '.com' suffix to their name, as a means of signalling that they were capitalising on the internet boom. The network effects and first-mover advantage associated with internet ventures led most dot-com firms to adopt a 'get big fast' business model (Olivia, Sterman and Giese 2003) that prioritised growth over profits.

During this period, there was a huge inflow of capital into the stock markets that sent all major stock indices surging. In particular, the tech-heavy Nasdaq Composite rose from 549.30 points on January $2^{nd}$ 1990 to a historical high of 5048.62 points on March $10^{th}$ 2000. Alas, the Nasdaq did not reach its peak again until $23^{rd}$ April 2015; an immense crash set in that sent the index down to a trough of 1108.49 points on $10^{th}$ October 2002 (Source: CRSP).

Ultimately, many dot-com firms were bankrupted as venture capital dried up, Pets.com and Webvan.com being notable examples; Berlin (2008) provides an estimate that 52% of dot-com firms founded since 1996 did not survive past through 2004. Valuations fell massively, and many dot-coms were forced to revise accounts or write down goodwill.

The meteoric rise of the Nasdaq and its subsequent plunge have led many to question whether stocks were rationally valued at the time. The standard definition of an economic bubble is when the equilibrium price of an asset greatly exceeds its fundamental value, as determined by its expected cash flows and risk; it is not useful to characterise a period as a bubble purely because of a rise and fall in prices (Damodaran 2014). Thus, to identify bubbles, it is necessary to consider both the price action and the underlying fundamentals.

On the one hand, Cooper et al (2001) find that companies which changed their name to include a dot-com suffix experienced permanent increases in valuation, regardless of their actual level of involvement with the internet. Lamont and Thaler (2003) highlight six cases of indisputable mispricing where a subsidiary's valuation exceeded that of its parent company, such as the case of 3Com and Palm.

On the other hand, Pástor and Veronesi (2003, 2006) argue that high valuations may have been justifiable due to uncertainty about future profitability- there is always a slim chance that a company becomes 'the next Microsoft'- and that the so-called bubble burst when realised profitability turned out to be low. Lettau et al (2008) also suggest the high stock prices could be rationalised by a fall in macroeconomic risk and consumption volatility.

In general, there is a strong case for individual cases of irrational valuation. However, whether the broader market was overvalued at the time is not particularly clear from the literature above. A number of papers have thus risen to answer this question from an econometric time series perspective, using a specification known as the rational bubble model. The rational bubble model has clear implications for the presence of a bubble component, but testing for these properties is non-trivial. There are a number of pitfalls, most notably the possibility of periodically collapsing bubbles which may manifest only temporary explosive behaviour. Indeed, Cuñado et al (2005) are the first to attempt to test for rational bubbles in the Nasdaq Composite but fail to account for this; they reject the presence of a bubble precisely because their test of mean reversion is intended to find persistent as opposed to temporary deviations from fundamentals.

Phillips, Wu and Yu's (PWY) 2011 paper thus marks a milestone in both the bubble-testing and dot-com bubble literature. They develop a test for rational bubbles that not only has the power to detect periodically collapsing bubbles, but is also able to identify when the bubble started and ended. They apply their test to the Nasdaq Composite over the period 1973-2005 and find evidence for a bubble; they identify its start date as July 1995, with the end date lying somewhere between September 2000 and March 2001.

However, PWY's results also serve to reignite a divide in the literature that has largely gone unaddressed in recent years: whether the dot-com bubble began in 1995 or 1998. Indeed, Cuñado et al (2012) modify their original methodology and only find evidence for a bubble after June 1998 when using weekly data. It is worth noting that prior to PWY, many academic papers covering the dot-com bubble chose to focus on the period 1998-2000, implicitly defining this as the duration of the bubble (e.g. Cooper et al 2001, Ofek and Richardson 2003). PWY themselves do not mention this; to my knowledge, only two sources from the mid-2000s explicitly acknowledge this divide.

The first source is *Manias, Panics and Crashes* by Aliber and Kindleberger (2005). They posit that there are two potential starting dates for the dot-com bubble: spring 1995 and summer 1998. They attribute the former's cause to the 1994 Mexican peso crisis, where the depreciation of the peso greatly increased the US trade deficit with Mexico and thus brought in vast capital inflows. The latter is attributed to the depreciation of Asian currencies following the mid-1997 Asian financial crisis, which likewise led to a large US trade deficit and generated capital inflows; Aliber and Kindleberger argue that stock prices were subsequently pushed up even further by lower interest rates in response to the collapse of LTCM in summer 1998.

The second source is DeLong and Magin (2006), who highlight that in the aftermath of the dot-com bubble, most commentators recognised the start of bubble as being somewhere between the Netscape IPO in August 1995 and Alan Greenspan's irrational exuberance speech in December 1996. However, Delong and Magin argue that based on implied excess stock returns, the Nasdaq was not experiencing a bubble until sometime between April 1997 and October 1998. They suggest that prior exuberance was not irrational, because investing in the Nasdaq in August 1995 would have earned an average real return of 9.3% annually through to November 2005, despite the bursting of the bubble in between.

Unfortunately, neither source is conclusive. Aliber and Kindleberger rely solely on qualitative analysis, whilst DeLong and Magin rely on heuristic measures of required stock returns. Schwert (2002) does implicitly provide a more quantitative characterisation of the divide: using a GARCH model, he finds that Nasdaq volatility rose significantly after 1998. Then, comparing value-weighted and equal-weighted Nasdaq portfolios, he pinpoints a significant rise in volatility after mid-1998 that was driven by technology stocks. This result is consistent with Cuñado et al (2012), who find a significant structural break in 1998 'for all model specifications and data periodicity'. Yet, Schwert also finds that the Nasdaq briefly exhibited unusual volatility relative to the S&P Composite in the latter half of 1995; the relative level of volatility falls in 1996 and remains stable until mid-1999.

Whether the dot-com bubble began in 1995 or 1998 has important implications. DeLong and Magin highlight that historically, bubble episodes have typically been very short, lasting no more than a couple of years. A start date of 1995 would imply the dot-com bubble lasted more than triple the historical average duration: this raises the question of whether the dot-com bubble was substantially different in nature to other bubbles, or, if it were a bubble at all in the first place.

Thus, I propose to reconcile this divide by testing two different explanations: 1) that the dot-com bubble was not one continuous entity, but instead comprised multiple distinct periods of exuberance; and 2), that exuberant stock prices were initially justified by fundamentals, and did not constitute a bubble until later on. Note that these two explanations do not contradict each other and can hold simultaneously.

I argue that PWY are unable to account for the first explanation because they apply their bubble test to the Nasdaq Composite, which is an aggregated value-weighted series containing most stocks listed on the Nasdaq Exchange. Whilst the Nasdaq Composite provides a good measure of overall capital inflows, the aggregation process completely conceals the heterogeneity amongst the underlying series- there are many non-tech stocks listed on the Nasdaq as well. Indeed, Pavlidis et al (2017) show that testing for bubbles in aggregated series can lead to significant power losses: temporary explosiveness in individual series may be obfuscated, and opposing dynamics may cancel each other out once aggregated.

PWY are also unable to rigorously account for the second explanation because they use the Nasdaq Composite's dividend series as a measure of fundamentals. However, this is not representative of the underlying series' fundamentals at all: 72% of all stocks within my sample did not pay dividends at all, which increases to 91% for information technology stocks. PWY effectively treat such stocks as having zero fundamental value within their aggregate specification, despite the fact that non-dividend-paying stocks ought to be valued on the basis of potential or expected future dividends. This means that PWY's approach is biased towards finding a bubble: to find robust deviations from a fundamental value requires a robust fundamental specification in the first place.

To address the first issue, I decide to adopt the most direct approach: instead of testing the aggregate series for bubbles, I test each of the individual underlying series then aggregate the results. To address the second issue, I consider alternate fundamental specifications instead of dividends. However, I do follow PWY in using the rational bubble framework and choose to use the extension of their tests as developed by PSY, both of which are outlined in the next section.

## 3. Theory

It is first worth noting that a number of other models have been applied to the dot-com bubble. For instance, Ofek and Richardson (2003) and Hong et al (2006) both explore models in which bubbles arise due to heterogeneous beliefs and short-sales constraints; such models predict that bubbles should shrink when faced with lockup expirations and insider selling, which they argue is consistent with the dot-com bubble circa Winter 2000.

However, whilst such models are able to capture complex agent dynamics and can be calibrated with data to help estimate certain properties of bubbles, they do not provide a clear bubble identification strategy. In contrast, the standard rational bubble model eschews more complex dynamics by assuming homogeneous agents and symmetric information, but forms a highly tractable basis for testing bubbles. It is the most widely used model in the bubble testing literature and will also serve as the core model for this paper.

The central concept underpinning the rational bubble model is that the current price of an asset depends not only on its expected fundamental returns (e.g. dividends, rent) but also its expected price in the next period. Rational agents may be willing to buy an asset priced above its fundamental value if the asset is expected to have an even higher selling price in the next period, and thus those expectations become self-fulfilling. As such, despite commonly being associated with irrationality, bubbles can arise even amongst agents with rational expectations, hence the name of the model, which can be derived as follows:

Suppose a riskless bond with net zero supply is available as an alternative investment, which yields a constant net interest rate of $r$ each period. Then, in equilibrium, no arbitrage implies that the expected stock return $R$ is equal to the risk-free rate of return $1 + r$, where $P_t$ is the asset price and $D_t$ is the asset payoff.

$$R = \frac{E_t(P_{t+i} + D_{t+i})}{E_t(P_{t+i-1})} = 1 + r$$

Setting $i = 1$ and taking rational expectations yields the following difference equation:

$$E_t(P_t) = P_t = \frac{1}{1+r} E_t(P_{t+1} + D_{t+1}) \tag{1}$$

Equation (1) can then be iterated forward using the law of iterated expectations. If we let $T$ denote the total lifetime of the asset, then solving forward $T - t - 1$ times yields:

$$P_t = \sum_{j=1}^{T-t} \frac{E_t(D_{t+j})}{(1+r)^j} + \frac{E_t(P_T)}{(1+r)^{T-t}} = P_t^f + B_t \qquad (2)$$

Where the two terms on the RHS of (2) are denoted as the fundamental component $P_t^f$ and the bubble component $B_t$ respectively. For a finite maturity asset like a bond, the terminal price $P_T = 0$. Given that agents have rational expectations, the price in the penultimate period is equal to purely its fundamental value, and by backward induction, this applies to all earlier prices. Rational bubbles therefore do not arise under finite horizons.

However, this is not necessarily the case for assets with infinite maturity, such as stocks. Under infinite horizons, the stock price $P_t$ will only equal its fundamental value (the infinite sum of expected future discounted dividends, otherwise known as the dividend discount model) if the bubble component converges to zero: $\lim_{T \to \infty} B_t = \lim_{T \to \infty} \frac{E_t(P_T)}{(1+r)^{T-t}} = 0$. This is known as the transversality condition. If this doesn't hold, then any $B_t$ that satisfies $B_t = E_t\left[\frac{B_{t+1}}{1+r}\right]$ is a solution to the difference equation, not just $B_t = 0$. This demonstrates a key property of rational bubbles: they grow explosively in expectation at the net rate $r$, and are thus submartingales.

Testing for rational bubbles is riddled with pitfalls: PWY's test represents the culmination of many lessons across decades of research. Firstly, the bubble component cannot be estimated directly in a regression because it is explosive: the central limit theorem no longer holds and inference is not robust (Flood and Hodrick, 1990). Secondly, because there exists a multiplicity of solutions to the bubble component, one should look to reject the 'no-bubble' hypothesis instead of testing for a specific form of a bubble (Blanchard and Watson 1982). Early candidates for indirect bubble tests included the variance bound tests by Shiller (1981) and LeRoy and Porter (1981), but when modified to relax stationarity assumptions and reduce small sample bias (Mankiw et al 1985), Gürkaynak (2008) highlights that a violated variance bound no longer has any bubble implications at all. Another indirect test is West (1987)'s two-step Hausman specification test, which compares two estimates of the discount rate. West generates two estimates of the discount rate from equations (1) and (2), one which is consistent regardless of a bubble and another which is only consistent if there is no bubble. The problem with West's test is that it is extremely sensitive to the fundamental model being correctly specified, such as the autoregressive dividend process; it would not be practically feasible to check all misspecification possibilities.

There is thus the need for indirect bubble tests that do not require the fundamental specification to be as exact. This precisely the purpose of Diba and Grossman (1988b)'s integration and cointegration tests, which allow for unobserved additive fundamentals provided they are stationary. Diba and Grossman's tests rely on three results:

The first is that stock prices and dividends are cointegrated of order (1,1) within the net present value model, assuming that both are I(1). This can be shown by rearranging equation (2) to give:

$$P_t - \frac{D_t}{r} = B_t + \frac{1}{r}\left[\sum_{j=1}^{\infty}\frac{E_t(\Delta D_{t+j})}{(1+r)^{j-1}}\right] \tag{3}$$

If $\Delta D_{t+j}$ is stationary and there are no bubbles, then $P_t - \frac{D_t}{r}$ is stationary. Hence, prices and dividends are cointegrated with the vector $(1, -\frac{1}{r})$.

The second result Diba and Grossman use is that in expectation, an explosive bubble component remains non-stationary no matter how many times it is differenced. The third result comes from Diba and Grossman (1987, 1988a), who show that if a rational bubble is present, it must have existed since the start of trading and cannot restart after fully collapsing. The third result implies that the second result holds not just in expectation, but in realisation as well (Gürkaynak 2008). The collective implication is that if there is a bubble, its explosive nature would render the stock price more non-stationary than the dividend series. In that case, if dividends need to be differenced N times to be stationary, stock prices would still be non-stationary after being differenced N times. Note that explosive stock prices could be driven by explosive dividends, so it is crucial to check the dividend process.

Diba and Grossman thus use Dickey-Fuller unit root tests, Bhargava cointegration tests and autocorrelation patterns to test these implications. They find evidence to suggest that prices and dividends are both I(1) and cointegrated, thus rejecting the presence of a bubble in their data.

However, Evans (1991) points out a critical flaw in the third result that Diba and Grossman use. He highlights that a bubble may collapse to a small non-zero manner before restarting, such that the second result they use may no longer hold in realisation, greatly reducing the power of their tests. This is because the periodically collapsing nature of Evans' bubble manifests temporary explosive behaviour that can cause the stock price to appear more like an I(1) or stationary series.

Given that conventional unit root and cointegration tests cannot detect periodically collapsing bubbles, PWY (2011) therefore propose the supremum augmented Dickey-Fuller test (SADF). The SADF uses the standard ADF regression equation but with asymptotically negligible drift. It is a right-tailed test such that the alternative hypothesis is explosive:

$$y_t = dT^{-\eta} + \beta y_{t-1} + \sum_{i=1}^{k} \delta_i \Delta y_{t-i} + \varepsilon_t , \ \ \varepsilon_t \sim^{iid} (0, \sigma^2) \tag{4}$$

$$H_0: \beta = 1 \ (unit \ root)$$
$$H_1: \beta > 1 \ (explosive)$$

Where $y_t$ is the time series being tested, $k$ represents the number of lags, $d$ is a constant, $T$ is the sample size and $\eta$ is a coefficient controlling the magnitude of the drift. Unlike the conventional ADF, the SADF test statistic is estimated recursively over different subsamples of the data and is therefore more sensitive to temporary explosive behaviour. Specifically, let $r_1$ and $r_2$ denote the start and end points of each subsample, where they represent fractions of the whole sample $T$ (hence $r_{2 \ max} = 1$). Then, let the minimum subsample window size be $r_0$ and fix $r_1$ at 0, whilst increasing $r_2$ from $r_0$ to 1, allowing the subsample size to be incremented by one observation each time. The above ADF regression is estimated for each subsample and an ADF test statistic is generated. The SADF test statistic is then calculated as the supremum of all the subsample ADF test statistics:

$$SADF(r_0) = \sup_{r_2 \in [r_0, 1]} ADF_0^{r_2} = \sup_{r_2 \in [r_0, 1]} \frac{\int_0^{r_2} \widetilde{W} dW}{\left( \int_0^{r_2} \widetilde{W}^2 \right)^{\frac{1}{2}}} \tag{5}$$

Where $W$ is the standard Brownian motion and $\widetilde{W}(r) = W(r) - \frac{1}{r} \int_0^1 W$ is the demeaned Brownian motion. The SADF test statistic can then be compared against its critical values, which are non-standard and obtained by Monte Carlo simulation. If the null is rejected in favour of an explosive unit root, explosive behaviour can then be date-stamped. To do this, the $ADF_0^{r_2}$ test statistics generated at each $r_2$ can be compared against the right-tailed critical values of the standard asymptotic Dickey-Fuller distribution (note that the standard ADF test uses the left-tailed critical values instead), which are also obtained by Monte Carlo simulation. The origination date is the $r_2$ corresponding to the first subsample for which the $ADF_0^{r_2}$ exceeds its critical value, whilst the collapse date is the first $r_2$ (post-origination date) for which $ADF_0^{r_2}$ falls below its critical value. Effectively, the origination date is marked by the marginal observation that causes the test statistic to fall into its critical region.

Homm and Breitung (2012) show that PWY's test has significantly greater power in detecting periodically collapsing bubbles than recursive modifications of structural break tests (e.g. Kim 2000). However, Phillips, Shi and Yu (2015) find that the SADF can be inconsistent when there are multiple episodes of exuberance and collapse. In particular, PWY's dating strategy uses all data from $r_1 = 0$ up to $r_2$ to calculate $ADF_0^{r_2}$, but because this data may contain multiple collapsing bubbles, it could result in finding pseudo-stationary behaviour.

PSY thus suggest using a more extensive range of subsamples by allowing a flexible subsample starting date $r_1$. They find that this yields substantially greater power than the SADF. This test is denoted as the generalised SADF (GSADF):

$$GSADF(r_0) = \sup_{\substack{r_2 \in [r_0, 1] \\ r_1 \in [0, r_2 - r_0]}} ADF_{r_1}^{r_2}$$

$$= \sup_{\substack{r_2 \in [r_0, 1] \\ r_1 \in [0, r_2 - r_0]}} \left\{ \frac{\frac{1}{2} r_w [W(r_2)^2 - W(r_1)^2 - r_w] - \int_{r_1}^{r_2} W(r) dr [W(r_2) - W(r_1)]}{r_w^{1/2} \left\{ r_w \int_{r_1}^{r_2} W(r)^2 dr - \left[ \int_{r_1}^{r_2} W(r) dr \right]^2 \right\}^{1/2}} \right\} \tag{6}$$

Where $r_w = r_2 - r_1$ and $W$ is the standard Wiener process. The GSADF critical values are obtained by Monte Carlo simulation and if the null is rejected, a modified date-stamping procedure can be executed via the backward SADF statistic:

$$BSADF_{r_2}(r_0) = \sup_{r_1 \in [0, r_2 - r_0]} ADF_{r_1}^{r_2}$$

The BSADF effectively takes the supremum of right-tailed ADF statistics calculated over backward-expanding samples, whose endpoint is fixed at a given $r_2$. The fractional start point $\hat{r}_e$ can then be determined by finding the $r_2$ corresponding to the first subsample for which the $BSADF_{r_2}$ exceeds the 95% critical value of the SADF statistic, $scv_{r_2}^{95}$. Similarly, the fractional end point $\hat{r}_f$ is the $r_2$ corresponding to the first subsample (post $\hat{r}_e$) for which the $BSADF_{r_2}$ falls below $scv_{r_2}^{95}$:

$$\hat{r}_e = \inf_{r_2 \in [\, r_0, 1]} \{r_2 : BSADF_{r_2}(r_0) > scv_{r_2}^{95}\}$$

$$\hat{r}_f = \inf_{r_2 \in [\, \hat{r}_e, 1]} \{r_2 : BSADF_{r_2}(r_0) < scv_{r_2}^{95}\}$$

The GSADF and BSADF tests have become dominant in the literature for their robustness and date-stamping ability. They have been applied to various asset markets, such as: Corbet et al (2018), who test bitcoin and ethereum; Hu and Oxley (2017), who test exchange rates; and Pavlidis et al (2018) who test crude oil. Brunnermeier et al (2020) further use PSY's test to identify bubble episodes that are then used as an independent variable in a regression explaining systemic risk.

I propose to apply the GSADF and BSADF to the individual stocks constituting the Nasdaq. If the stock price is explosive but the corresponding dividend series is not, then the stock is in a bubble and the start/end dates can be estimated. If both the stock price and dividend series are explosive, we cannot reject there being no bubble. However, 72% of the stocks in my sample do not pay dividends. Thus, I need to consider an alternate measure of fundamentals.

Caspi and Graham (2018) suggest testing the log book-to-market ratio (book value/market value of equity) for explosiveness if dividend data is not available. They use the following log-approximated specification derived by Vuolteenaho (1999):

$$\log\left(\frac{V_t}{M_t}\right) = k_t + E_t\left[\sum_{j=1}^{\infty} \rho^j r_{t+j+1}\right] - E_t\left[\sum_{j=1}^{\infty} \rho^j \left(r_{t+j+1}^e - r_{t+j+1}^f\right)\right] + B_t = P_t^f + B_t$$

Where $\frac{V_t}{M_t}$ is the book-to-market ratio, $r_t$ is the log gross extra return over the market, $r_t^e$ is the log gross return on equity, $r_t^f$ is the log gross risk-free return and $B_t$ is the bubble component. However, I argue there are two major problems with their approach. The first is that Vuolteenaho derives his specification by assuming market cap and book value are cointegrated in the first place. The second is that in their empirical application, Caspi and Graham ignore $P_t^f$ entirely and effectively assume it is stationary. Whilst the standard rational bubble model also assumes constant discount rates, the return on equity is calculated as net income/book value: it seems highly unlikely that net income is always stationary, and I indeed find evidence of explosive earnings in my data.

On the other hand, Basse et al (2021) argue that the smoothing of dividend payments can lead to misleading fundamentals. They thus propose to construct a hypothetical dividend series based on net income (earnings) and average dividend payout ratios. However, their approach involves a great deal of estimation and is performed at the stock index level; one would need to be extremely careful when estimating payout ratios at the individual stock level.

My chosen approach is similar to that of Basse et al, but far more direct. I choose to substitute dividends with 'potential dividends' instead, as measured by Free Cash Flow to Equity (FCFE), which is defined as follows by Damodaran (2010):

$$\begin{aligned} FCFE = \ &Net\ Income - Capital\ Expenditures - Acquisitions \\ &+ Depreciation\ \&\ Amortisation - \Delta Noncash\ Working\ Capital \\ &+ Net\ Borrowing - Preferred\ Dividends \end{aligned} \tag{7}$$

FCFE is widely used by financial practitioners in discounted cash flow (DCF) models to value stocks. Bancel and Mittoo (2014) find that 80% of their survey respondents use DCF models, whilst Rutterford (2004) highlights that DCF models gained widespread traction in the 1980s-90s. FCFE can be substituted seamlessly into the net present value model in place of dividends, such that the cointegration and bubble results all still hold. It is a specification that holds in both theory and practice.

The underlying logic here is that the free cash flows to equity are what shareholders are entitled to. They represent the cash flow available to common shareholders once investments/acquisitions have been made, new debt has been issued, current debt has been repaid and preferred stockholders have been paid. Depreciation and amortisation are non-cash charges and are thus added back in. The working capital adjustment is made because of the accrual accounting system, which records revenues at the point of sale, as opposed to when cash actually exchanges hands. For instance, suppose there were an increase in accounts payable. The corresponding expenses would have already been recorded within the calculation of net income, but the corresponding cash would not have actually been deducted from the company's cash balance yet (Damodaran 2010).

Damodaran (2013) also highlights that financial firms cannot be valued using conventional FCFE because of the fundamental difference in the way they operate: e.g. reinvestment cannot be defined in terms of capital expenditures because they have none. Instead, he proposes:

$$FCFE_{Financial\ Service\ Firm} = Net\ Income - Reinvestment\ in\ Regulatory\ Capital \quad (8)$$

I do not have data on reinvestment in regulatory capital and it cannot be readily estimated. Therefore, I assume it is stationary in levels and treat it as an unobserved fundamental, as suggested by Diba and Grossman (1988). My measure for financial firms' FCFE is thus solely their net income, which, as Damodaran highlights, is used by many analysts anyway.

## 4. Data and Methodology

Due to the large amount of data involved, I use Pandas for data wrangling. In order to test thousands of different series individually and store the results, I use PyStata to invoke Stata directly from my Python environment. I use the Stata package 'radf' (Baum and Otero, 2020) to run the GSADF test; if the unit root null is rejected in favour of an explosive root, I then run the BSADF date-stamping procedure. The rolling subsample window size is calculated as $r_0 = 0.01 + 1.8/\sqrt{T}$, as suggested as by PSY. The number of lags is chosen as $int(4(T/100)^{0.25})$ as suggested by Schwert (1989) for unit root tests.

My data is obtained from three databases: CRSP, Compustat and CRSP-Compustat Merged (CCM). CCM links the former two, which provide price data and fundamental data respectively. Fundamental data is only available on a quarterly basis, and the earliest data I can find for certain variables such as capital expenditures is 1983Q2. Therefore, whilst PWY study the period February 1973-June 2005 on a monthly basis, I study the period June 1983-June 2005 on a quarterly basis. I use the CRSP Exchange Tool to identify *all* stocks listed at some point on the NASDAQ within my chosen period. Using PERMNO as a linking identifier, I then obtain the corresponding price and fundamental data from CCM. Where calendar quarter data is unavailable, I normalise the date on which the observation was recorded to a quarterly basis. I also back out quarterly values from variables such as capital expenditures which are provided on a year-to-date basis. The variables in my dataset are outlined in Table 1 below:

*Table 1*

| Core Variables | Fundamental variables used to calculate FCFE (all measured in USD millions) | Global Industry Classification Standard (GICS) Variables |
|---|---|---|
| PERMNO (database identifier) | Net Income | Sector |
| Company Name | Cash and Short Term Investments | Group |
| Recorded Date | Capital Expenditures | Industry |
| Calendar Quarter | Debt in Current Liabilities | Subindustry |
| Price | Total Long Term Debt | |
| Common Shares Outstanding | Acquisitions | |
| Dividends per Share | Working Capital | |
| Stock Exchange | Depreciation and Amortisation | |
| | Preferred Dividends | |
| | Long Term Debt Issuance | |
| | Long Term Debt Reduction | |

Since this paper focuses on the NASDAQ, I need to exclude periods in which a stock was trading elsewhere. The problem is that the CCM stock exchange variable only refers to the exchange on which the stock was *last* traded, across all periods. For instance, Oracle's exchange is listed as the NYSE, despite the fact that it was traded on the NASDAQ until 2013. To remedy this, I input the list of PERMNOs into CRSP, since the exchange variable on CRSP refers to the exchange *at the time*- however, CRSP does not provide any data on fundamentals. Hence, I cross-check each stock series from my CCM data with the corresponding series from my CRSP dataset and exclude periods based on the CRSP exchange variable. This process reduces the number of observations from 400,139 to 339,707.

I also check each stock price series for gaps: perhaps the observation is missing, or perhaps the stock was delisted. Interpolation would not be suitable here since the gaps typically exceed one period, and delisting implies a structural change to the underlying company that results in vastly different prices and fundamentals after the break. Instead, I opt to split a series whenever there is a gap, and treat them effectively as separate units. I drop all series with fewer than 9 observations because at least 9 observations are required to run the BSADF. This reduces the sample to 328,514 observations, corresponding to 10,704 unique stocks (comprising 11,354 disjoint time series). This forms my base dataset, for which Table 2 provides summary statistics of net income and dividend payments, broken down by sector. Utilities, financials and real estate are the only sectors in which more than half of stocks pay dividends. The largest quarterly earnings within my sample were reported by Oracle in 2002Q2, whilst the largest quarterly loss was reported by Viavi Solutions in 2001Q2; the latter was previously known as JDSU and recorded the largest goodwill write-down ever at the time (Norris 2001).

*Table 2*

| Sector | Total number of Stocks | Non-dividend paying stocks | % of stocks that don't pay dividends | Net Income (mean) | Net Income (std) | Net Income (min) | Net Income (median) | Net Income (max) |
|---|---|---|---|---|---|---|---|---|
| *Utilities* | 80 | 27 | 33.75 | 1.48 | 10.80 | -535.73 | 0.73 | 32.26 |
| *Financials* | 1931 | 477 | 24.70 | 4.98 | 22.43 | -2435.15 | 1.40 | 581.09 |
| *Energy* | 428 | 355 | 82.94 | 1.73 | 35.27 | -294.71 | 0.00 | 1414.00 |
| *Information Technology* | 2391 | 2178 | 91.09 | -0.20 | 190.44 | -41847.90 | 0.12 | 4912.41 |
| *Healthcare* | 1428 | 1305 | 91.39 | -0.58 | 24.20 | -2601.60 | -0.16 | 1029.00 |
| *Industrials* | 1418 | 1053 | 74.26 | 1.46 | 23.87 | -903.00 | 0.27 | 3355.29 |
| *Consumer Discretionary* | 1845 | 1481 | 80.27 | 0.18 | 44.25 | -4235.68 | 0.24 | 2757.89 |
| *Materials* | 434 | 294 | 67.74 | 4.24 | 49.29 | -418.25 | 0.09 | 2678.00 |
| *Consumer Staples* | 368 | 240 | 65.22 | 2.68 | 16.55 | -301.60 | 0.42 | 526.39 |
| *Communication Services* | 285 | 243 | 85.26 | -8.08 | 164.57 | -5426.30 | -0.45 | 3176.00 |
| *Real Estate* | 41 | 19 | 46.34 | -0.71 | 8.45 | -143.91 | 0.18 | 60.72 |
| *Other* | 55 | 46 | 83.64 | -0.69 | 6.62 | -65.67 | -0.06 | 24.53 |
| *All* | 10704 | 7718 | 72.10 | 1.05 | 99.16 | -414847.90 | 0.25 | 4912.41 |

To calculate FCFE for non-financials, I follow equation (7); for financials, I just use net income. Specifically, I calculate the change in non-cash working capital as the following:

$$\Delta(working\ capital\ -\ cash\ and\ short\ term\ investments\ +$$
$$debt\ in\ current\ liabilities)$$

To calculate net borrowing, I consider two measures.

$$1.\ long\ term\ debt\ issuance\ -\ long\ term\ debt\ reduction$$
$$+\ \Delta debt\ in\ current\ liabilities$$
$$2.\Delta total\ long\ term\ debt\ +\ \Delta debt\ in\ current\ liabilities$$

In around 22% of observations these two definitions give exactly the same value. I suspect the disparity comes from the fact that long term debt issuance includes debt not classified into current and long term debt, as defined by the Compustat User's Guide. It is unclear in the wider literature as to which one is more appropriate to use: Hung et al (2017) use the former, whilst Dasgupta et al (2011) use the latter. I thus generate two separate estimates of FCFE: FCFE_1 and FCFE_2 respectively. A significant amount of data is missing for both: FCFE_1 has 203,084 observations spanning 7838 stocks, whilst FCFE_2 has 226,344 observations spanning 8356 stocks. Therefore, for robustness, I consider net income as an alternate fundamental specification for which 319,832 observations are available, spanning 10,557 stocks. I also consider normal dividends (62,258 observations spanning 1935 stocks) to provide a comparison with the conventional fundamental specification.

It is also important to note that unlike dividends, FCFE (and net income) can be negative. This can lead to econometric issues previously not encountered in the literature. Specifically, suppose FCFE is initially negative but experiences a brief period where it increases explosively, possibly becoming positive; intuitively, this should be a good sign for shareholders, who might increase their expectations of future cash flows. The problem here is that FCFE is modelled as an autoregressive process. Unless it continues to increase explosively for a sustained period through positive territory, it would instead be construed as a stationary or I(1) process, since it is effectively regressing towards zero. On the other hand, if a negative FCFE series were to *decrease* explosively, then this would be construed as autoregressive explosiveness by the GSADF.

To address this issue, I propose adding the constant $c = -FCFE_{min} + 1$ to each value in a FCFE series if FCFE is negative at any point. This ensures that the entire time series is positive. The cointegrating relationship between price and FCFE is preserved because it is specified in differences, which a constant transformation doesn't affect. To see this, consider adding a constant $-\frac{c}{r}$ to both sides of equation (3):

$$P_t - \frac{(D_t)}{r} - \frac{c}{r} = B_t + \frac{1}{r}\left[\sum_{j=1}^{\infty} \frac{E_t(\Delta D_{t+j})}{(1+r)^{j-1}}\right] - \frac{c}{r}$$

$$P_t - \frac{(D_t + c)}{r} = B_t + \frac{1}{r}\left[\sum_{j=1}^{\infty} \frac{E_t[(D_{t+j} + c) - (D_{t+j-1} + c)]}{(1+r)^{j-1}}\right] - \frac{c}{r}$$

The cointegrating relationship between $P_t$ and $D_t$ implies that $P_t$ and $D_t + c$ are cointegrated as well, since $-\frac{c}{r}$ is just a constant. Another way to see this is from Campbell and Shiller (1987)'s proof of cointegration: they show that $\Delta P_t$ can be expressed in terms of $\Delta D_t$, such that the properties of $\Delta D_t$ directly feed through if there are no bubbles. With a positive series, explosiveness in differences naturally translates into explosiveness in levels, but this is not the case for negative $D_t$, since explosiveness is considered on an autoregressive basis.

One might thus be tempted to test $\Delta P_t$ and $\Delta D_t$ for explosiveness instead, and avoid changing the properties of levels. The issue here is once again temporary explosiveness, which can cause a time series to appear like I(1) or even I(0) in levels, as previously mentioned by Evans (1991). Whilst the GSADF is able to detect this behaviour in levels, it would greatly struggle to detect any hints of temporary explosiveness if the I(1)/I(0)-like time series were differenced. Given that we need explosiveness in differences to translate to explosiveness in levels, but cannot just test differences, then adding a constant to the series in levels is the optimal solution.

## 5. Results and Robustness Checks

My criterion for identifying bubbles ('exuberance') is as follows: if there is evidence for explosiveness in the stock price, I then test the fundamental series for explosiveness. This is done by the GSADF. I then classify a stock as being in a bubble if the stock price is explosive in the current quarter, but there is no explosiveness in fundamentals in future periods, the current quarter or the previous quarter. This is accomplished via the BSADF date-stamping strategy.

This criterion is based on the notion that investors with rational expectations are forward-looking: upon receiving new information in the current period, they may revise their future expectations and trading strategies. The previous quarter's financial results are reported in the current quarter and are thus part of the current information set ($Fundamentals_{t-1} \in \Omega_t$), but not the previous quarter's information set.

In order to interpret results, Pavlidis et al (2016) suggest to run a panel extension of the GSADF test when testing individual series. However, this is not feasible here due to my highly unbalanced panel and large panel dimension. Purely looking at the number of stocks in exuberance at a given time would also be misleading, since exuberance in the stock of a large company might be overshadowed by exuberance in many smaller stocks. Therefore, I instead choose to aggregate the number of stocks in exuberance by sector and take the sum of their market cap for each period.

I obtain the following figures for FCFE_1, FCFE_2, dividends and net income respectively:

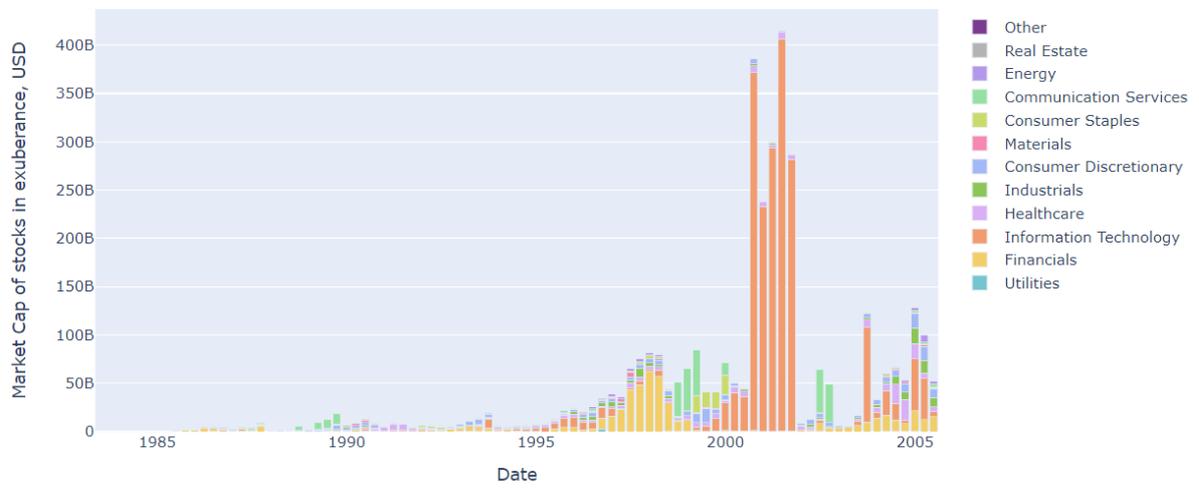

*Figure 1*

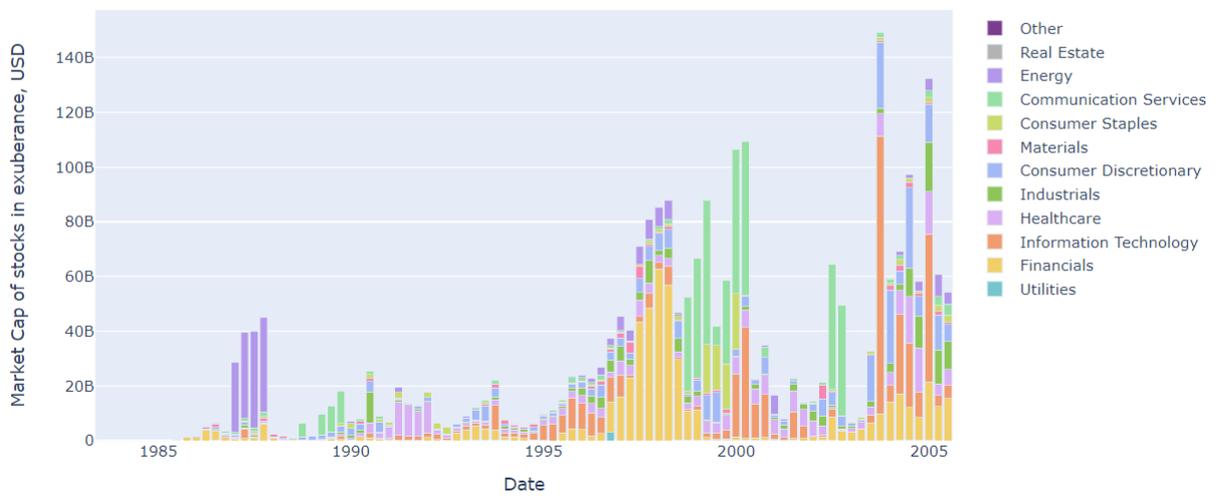

*Figure 2*

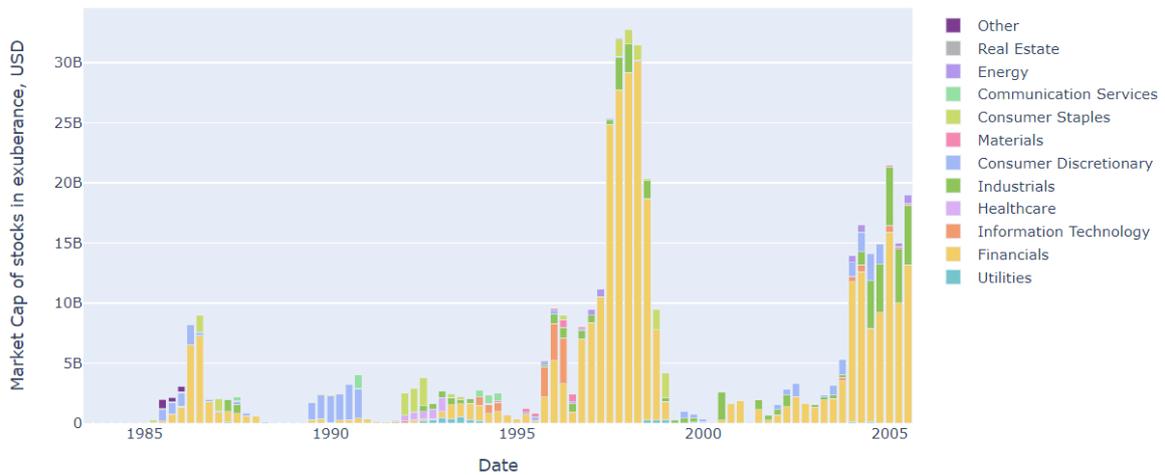

*Figure 3*

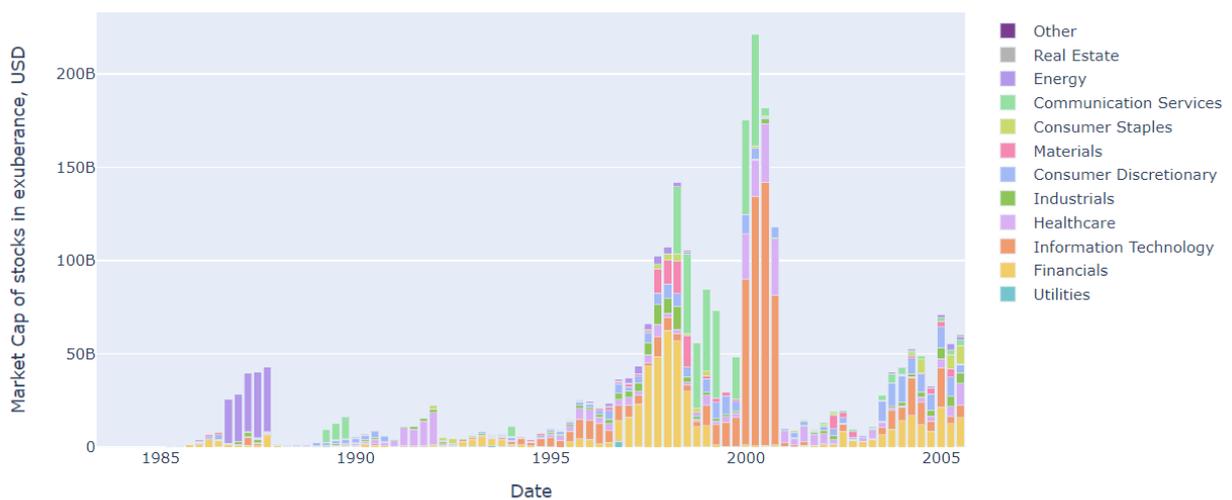

*Figure 4*

The key finding from these four figures is that the dot-com bubble was characterised by four overlapping episode of exuberance concentrated amongst three sectors. The first period of exuberance is in the information technology sector and begins around 1992, consistent with the acceleration in internet usage at the time. Starting from 1996, however, begins a substantial period of exuberance for financials, which lasts until 1998. This is most likely explained by the significant rise in IPOs following Netscape's resounding success in 1995, which would also have been bolstered by a policy change that led to an increase in the maximum amount of revenue banks could earn from underwriting (NYT 1996). The third period of exuberance is in the communications sector and lasts from the 1997 to 2000. The final episode is in the information technology sector again and lasts from 1998 until 2000. This is remarkably consistent with the hypothesis of there being two bubbles specifically for dot-com firms.

It is worth noting that the bubble criterion generally does not eliminate episodes of price exuberance for being fundamentally justified. Within the first three specifications, 861, 910 and 687 instances of price explosiveness respectively are not deemed as bubbles. However, in the case of the net income specification, 2855 instances are removed. The overall level of exuberant market cap is significantly lower. Figure 5 shows the original data for figure 4 before the net income bubble criterion was imposed: we see that this recovers the period of exuberance in information technology circa 1992Q4, which is seen in figures 1 and 2, but not 4. Notably, one of the companies constituting this period of exuberance is Intel Corp.

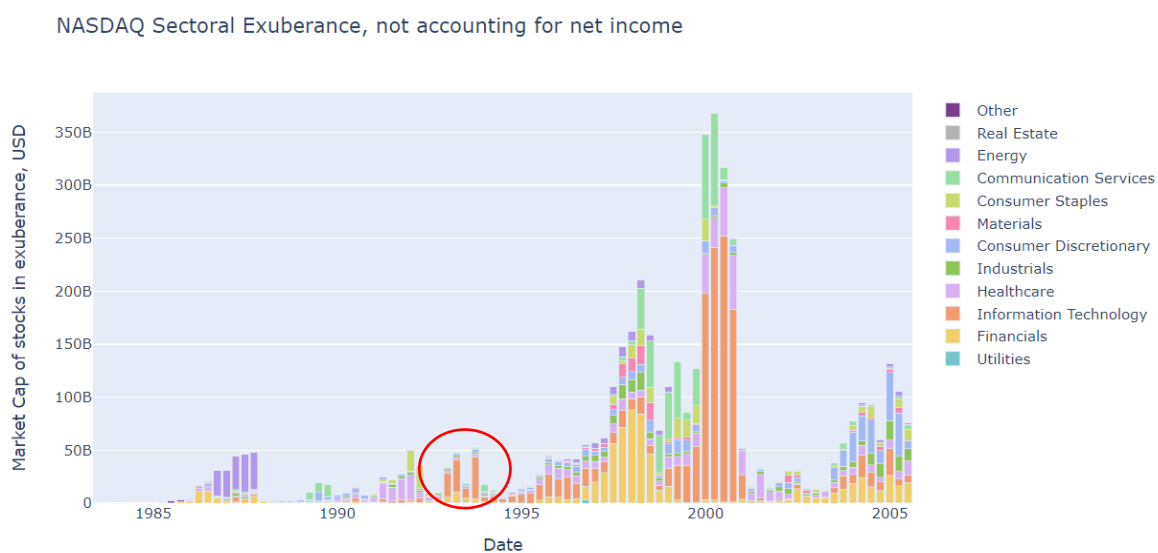

*Figure 5*

To estimate the dates of different episodes of exuberance more rigorously, I apply the GSADF and BSADF to the exuberant market cap data used in figure 5. Since the data is already 'exuberant', this is effectively a higher order moment: 'explosive exuberance'. It ought to capture the very peaks of exuberance within the underlying data. I find that exuberance in financials was explosive from 1997Q2 to 1997Q4, whilst exuberance in telecoms was explosive from 1998Q2 to 1998Q3, and then later for one quarter in 2000Q2.

For the information technology sector, I find that the first episode of exuberance lasted from 1992Q4 to 1993Q3 (which was effectively fundamentally justified by future net income), whilst the second lasted from 1999Q4 to June 2000. The key implication here is again that there were two episodes of internet exuberance, of which both were fundamentally justified to an extent, especially the former.

For subsamples of all the tests I run, I implement a number of robustness checks. Firstly, I experiment with the lag length in the ADF regression specification. I try using the AIC to select the lag length as suggested by Phillips and Yu (2011), and also the BIC as suggested by PSY (2015). I also try restricting the lag length to 1, as PSY highlight that the size of the test increases with the lag length. None yield significant differences.

Additionally, I try taking logs of the price and fundamental series (which are positive due to my constant transform), in accordance with Campbell and Shiller (1988) who derive a log-linear specification in which the two are cointegrated. This specification allows for a time-varying discount rate as long as it is stationary. Naturally, this specification is more parsimonious towards finding bubbles, but I find that the overall episodes of exuberance remain the same.

Furthermore, I implement the wild bootstrap as suggested by Phillips and Shi (2020) to mitigate heteroskedasticity. The wild bootstrap works by using the residuals from the initial regression to generate a bootstrap sample, from which bootstrapped GSADF test statistics are then calculated. I choose to use 200 replications and apply the wild bootstrap to FCFE_1; the computational intensity of the wild bootstrap combined with my large datasets makes it unsuitable for widespread use within the scope of this paper. The wild bootstrap is more parsimonious and I find that the anomalously high level of exuberance in information technology in figure 1 is slightly reduced.

Last but not least, I try testing all the price series from my monthly CRSP dataset, which does not include any fundamental data. This is because Monschang and Wilfling (2020) find that the BSADF is sensitive to data frequency. The results here are not directly comparable as the CRSP dataset contains a significantly larger panel, but the episodes of exuberance corroborate my original findings.

## 6. Conclusion

Overall, there is significant and robust evidence to support my first hypothesis: that the dot-com bubble was characterised by multiple episodes of exuberance across sectors, and that there were two potential starting dates- in fact, I find the dot-com bubble to have started as early as 1993 instead. There is also evidence to support my second hypothesis of the dot-com bubble being partially justified by fundamentals, but this is an important direction for future research. The joint hypothesis problem of differentiating between model specification and a bubble is hard to overcome.

In general, the level of heterogeneity I have uncovered from beneath an aggregate series strongly suggests that future papers on bubbles should also focus on testing for individual series. With more data available on sentiment and options, this is a promising area of research.